\documentstyle[11pt,newpasp,epsf,twoside]{article}
\def\edcomment#1{\iffalse\marginpar{\raggedright\sl#1\/}\else\relax\fi}
\reversemarginpar

\begin{document}
\title{Dwarf Nova Oscillations and Quasi-Periodic Oscillations: Extension of the Two-QPO
Diagram of X-Ray Binaries, and a new kind of DNO}

\author{Brian Warner and Patrick A. Woudt} 

\affil{Department of Astronomy, University of Cape Town, Rondebosch 7700, South Africa}

\begin{abstract}
Seventeen examples are given of Cataclysmic Variable (CV) stars possessing both 
Dwarf Nova Oscillations (DNOs) and Quasi-Periodic Oscillations (QPOs). These form an extension
of the X-Ray Two-QPO correlation to frequencies three orders of magnitude lower. We draw 
attention to the existence of a second type of DNO in CVs, which is probably caused by magnetically
channelled accretion onto the white dwarf.
\end{abstract}

\section{Introduction}

    We reported at the August 2001 G\"{o}ttingen meeting on CVs that VW Hyi in outburst has 
Dwarf Nova Oscillations (DNOs) and Quasi-Periodic Oscillations (QPOs) that keep 
an approximately constant period ratio of $\sim$ 16 as they vary in period by a factor 
of two during the late stages of outburst (Warner \& Woudt 2002a; see also Woudt \& Warner 2002). 
Interpreting the DNOs and QPOs in VW Hyi as analogues of the high and low frequency QPOs 
observed in X-Ray binaries, we showed that VW Hyi lies on an extension to low frequencies 
of the two-QPO correlation in X-Ray binaries found by Psaltis, Belloni \& van der Klis (1999). 
An important addition to this correlation was made by Mauche (2002), who found that the X-Ray 
DNOs and QPOs in SS Cyg also fall on this relationship, which helps to close the gap 
between the CVs and the X-Ray binaries.

    One aspect of QPOs in CVs is that double DNOs are sometimes observed, and their frequency 
difference is equal to the QPO frequency (Woudt \& Warner 2002) - which we interpret as the DNO 
rotating beam reprocessed off a progradely travelling wave in the inner disc (Warner \& Woudt 2002b). 
Therefore, even if a QPO is not directly observed in a light curve, a proxy for it can 
be found when double DNOs are observed.

\vfill\eject

\section{The Two-QPO Diagram for CVs and X-Ray Binaries}

    Armed with this knowledge we have searched the literature for examples of the 
simultaneous appearance of DNOs and QPOs, or their proxies. Table 1 lists the results of the search.

\begin{table}
 \centering
  \caption{DNOs and QPOs in CVs (periods in seconds).}
  \begin{tabular}{l r r l}
\\[2pt]
\hline \\ \vspace{1mm}
Star & DNOs & QPOs & References \\[5pt]
\hline \hline \\[1pt] 
V3885 Sgr & 29.08 \hfill 30.1  & [819] & Hesser, Lasker \& Osmer 1974\\
WZ Sge    & 27.87 \hfill 28.95 &  742  & Warner \& Woudt 2002b \\
SS Cyg    &  7.7        &   83  & Mauche 2002 \\
V436 Cen  & 19.6        & 475   & Warner \& Woudt 2002b\\
          & 19.45 \hfill 20.29 & [470] & Warner \& Woudt (unpublished)\\
V2051 Oph & 28.06 \hfill 29.77 & [489] & Steeghs et al. 2001\\
UX UMa    & 29          & 650   & Nather \& Robinson 1974\\
IX Vel    & $\sim$ 28   & $\sim$ 500 & Warner et al. 1985\\
          &             &            & Williams \& Hiltner 1984\\
TY PsA    & 27          &  245  & Warner et al. 1989\\
GK Per    & 351         & $\sim$ 5000 & Morales-Rueda et al. 1996, 1999 \ \ \ \ \\
RX And    & 35.7        & $\sim$ 1200 & Szkody 1976\\
SW UMa    & 22.3        & $\sim$ 300 & Robinson et al. 1987\\
RU Peg    & 2.94?       & 50 & Patterson et al. 1977\\[5pt]
\hline
\end{tabular}
\label{tab1}
\newline
{\small{Notes: values in square brackets are deduced from double DNOs. 
In RU Peg the observed DNO
period is 11.1 s, which is probably a beat with the integration length.}}
\end{table}

    In addition, we have four CVs in which we have recently observed DNOs and QPOs; these are listed in Table 2.

\begin{table}
 \centering
  \caption{Newly observed DNOs and QPOs (periods in seconds).}
  \begin{tabular}{l l r r}
\\[2pt]
\hline \\ \vspace{1mm}
Star & Type & DNOs & QPOs \\[5pt]
\hline \hline \\[1pt] 
WX Hyi   & Dwarf Nova & 12.1 & 190 \\
VZ Pyx       &   Dwarf Nova        &     23.86       &            390.5 \\
EC 2117-54  &  Nova-like    & 22.11 \ \ \hfill  23.28          &          470\\
HX Peg      &     Dwarf Nova   &           112.28   &               1800\\[5pt]
\hline
\end{tabular}
\label{tab2}
\end{table}

\begin{figure}[t]
\plotfiddle{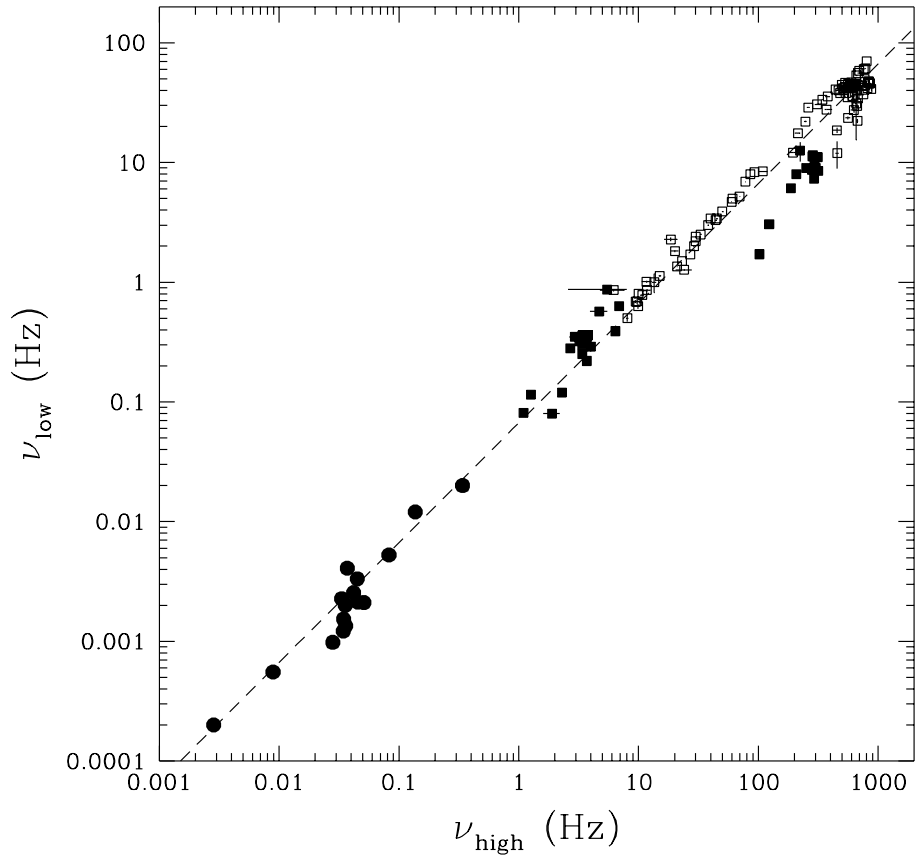}{11.5cm}{0}{65}{65}{-205}{-125}
\caption{The Two-QPO diagram for X-Ray Binaries (filled squares: black hole binaries; open squares: neutron star binaries)  
and 17 CVs (filled circles). The X-Ray binary data are from Belloni, Psaltis \& van
der Klis (2002) and were kindly provided by T. Belloni. The dashed line marks P$_{QPO}$/P$_{DNO}$ = 15.}
\label{warnerfig1}
\end{figure}

       The addition of these data, obtained from observing the literature and observing 
the sky, to the Two-QPO diagram produces the correlation shown in Figure 1. 
It can be seen that the CV DNO and QPO observations lie on an extension of the X-Ray 
Binary relationship to frequencies nearly three orders of magnitude lower. 
The essence of both the CV and the X-Ray correlations is that P$_{QPO}$ $\sim$ 15 P$_{DNO}$.

    Adopting a magnetically controlled accretion model for DNOs (Paczynski 1978; Warner 1995; 
Warner \& Woudt 2002b), the DNO period is clearly related to the Keplerian period at the 
magnetically truncated inner edge of the disc.  A number of recent papers modelling 
accretion into magnetospheres find quasi-periodic accretion or oscillations 
(e.g. Uzdensky 2002; Titarchuk \& Wood 2002); a quantitative model by Goodson, Bohm \& Winglee 
(1999) deduces P$_{QPO}$/P$_{DNO}$ $\sim$ $100/2{\pi} \sim 16$.

\section{A New kind of DNO}

     During the literature search for DNOs and QPOs we noticed that there are occasionally 
additional DNOs listed that do not fall into the above category - indeed, several CVs show 
two sets of DNOs simultaneously. Furthermore, the second kind of DNOs behave very differently 
from the `classical' DNOs, in that their periods are relatively independent of 
outburst state, and they can appear during quiescence of dwarf novae (which, apart from 
the apparently more strongly magnetic systems WZ Sge and GK Per, normal DNOs do not).

    The new type of DNOs, which we will refer temporarily as `longer period DNOs' -  
lpDNOs, are demonstrated in the following stars: \\

{\bf VW Hyi:} In a superoutburst of VW Hyi observed in 1975 Haefner, Schoembs \& Vogt (1977, 1979, 
see also Schoembs 1977) found an $\sim$ 80 s DNO which persisted for nine nights and changed 
only slightly and erratically in period although the star decreased in brightness from mag 
10.6 to 13.7. We have also seen occasional DNOs near 88 s in VW Hyi. The normal DNO periods 
in VW Hyi range from 14 s at maximum of outburst to 40 s near return to quiescence (Woudt \& Warner 2002).

{\bf EC 2117-54:} This newly discovered nova-like CV, from the Edinburgh-Cape Survey, 
possesses normal (including double) DNOs near 23 s and QPOs near 450 s, but in addition 
frequently shows lpDNOs near 94 s (see Figure 2). Thus EC 2117-54 has a suite of 
oscillations that closely resembles that of VW Hyi in outburst.

\begin{figure}
\plotfiddle{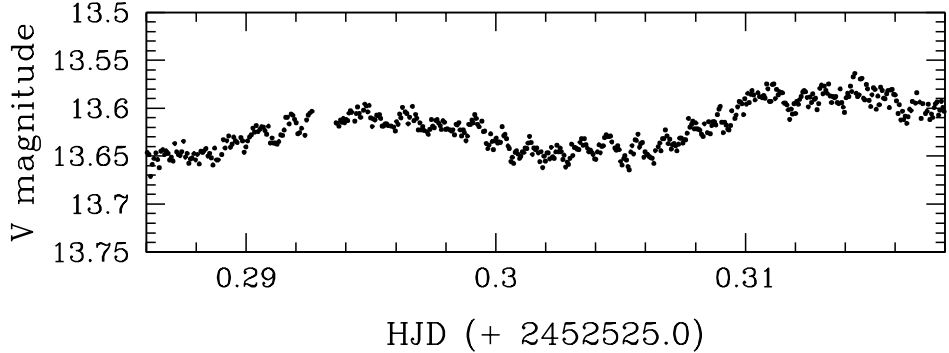}{4.5cm}{0}{86}{86}{-215}{-445}
\caption{Longer period DNOs observed in EC 2117-54}
\label{warnerfig2}
\end{figure}

{\bf SS Cyg:} Robinson \& Nather (1979) and Patterson (1981) observed what they described 
as QPOs in the range 32 -- 36 s in two different outbursts of SS Cyg, at the same 
time  $\sim$ 10 s normal DNOs were present.

{\bf HT Cas and AH Her:}   Patterson (1981) observed $\sim$ 100 s QPOs in HT Cas and 
AH Her in outburst, where the normal DNOs are $\sim$ 20 s and $\sim$ 25 s respectively. 
The HT Cas oscillations were also often seen during quiescence.

{\bf OY Car:}  We have observed strong $\sim$ 50 s oscillations during one observational 
run on OY Car at quiescence. This star has normal DNOs of 18 -- 28 s during outbursts.

   For a possible interpretation of the lpDNOs we turn to measurements of $v \sin i$ of the white 
dwarf primaries. In VW Hyi Sion et al. (1996, 2001) have found projected rotational velocities 
in the range 300 -- 500 km s$^{-1}$. For $i$ = 65$^\circ$ and a primary mass of 0.6 M$_\odot$ 
this gives a rotation period of 120 -- 165 s. The observed lpDNO may therefore be associated 
with the bulk rotation of the primary (as opposed to the rotation of the equatorial belt 
in the case of the normal DNOs - Warner \& Woudt 2002a,b), with two-pole magnetic accretion. 
A similar result is obtained for SS Cyg, with $v \sin i$ = 300 km s$^{-1}$ (Sion 1999), $i$ = 50$^\circ$ 
and M = 1.2 M$_\odot$, which give a primary rotation period $\sim$ 63 s.

   We suggest, therefore, that the lpDNOs result from accretion controlled by the 
magnetic field of the primary itself, whereas the normal DNOs come from accretion onto 
the equatorial belt where a stronger field is generated by differential rotation.

\bigskip
\acknowledgements{BW is supported by funds from the University of Cape Town; 
PAW is supported by strategic funds made available to BW by the University of 
Cape Town and by the National Research Foundation. We thank T. Belloni for kindly providing
us with the X-Ray Binary data shown in Figure 1.}

\end{document}